\documentclass[12pt]{iopart}
\usepackage{cite}
\usepackage[dvips]{graphicx}
\renewcommand{\lsim}{~{\buildrel < \over {_\sim}}~}
\newcommand{\gsim}{~{\buildrel > \over {_\sim}}~}
\newcommand{\sqrtsNN}{\sqrt{s_{\scriptscriptstyle{{\rm NN}}}}}

\newcommand{\mev}{\mathrm{MeV}}
\newcommand{\gev}{\mathrm{GeV}}
\newcommand{\tev}{\mathrm{TeV}}
\newcommand{\fm}{\mathrm{fm}}

\newcommand{\mum}{\mathrm{\mu m}}

\newcommand{\PbPb}{\mbox{Pb--Pb}\ }
\newcommand{\pPb}{\mbox{p--Pb}\ }

\newcommand{\NN}{\mbox{nucleon--nucleon}}

\renewcommand{\AA}{\mbox{nucleus--nucleus}}
\newcommand{\RAA}{R_{\rm AA}}

\renewcommand{\pt}{p_{\rm t}}
\renewcommand{\d}{{\rm d}}
\newcommand{\dEdx}{{\rm d}E/{\rm d}x}
\newcommand{\dNdy}{{\rm d}N_{\rm ch}/{\rm d}y}

\newcommand{\QQbar}{\mbox{$\mathrm {Q\overline{Q}}$}}

\newcommand{\ccbar}{\mbox{$\mathrm {c\overline{c}}$}}
\newcommand{\bbbar}{\mbox{$\mathrm {b\overline{b}}$}}

\newcommand{\Dz}{\mbox{$\mathrm {D^0}$}}

\begin{document}

\title{Charm and beauty of the Large Hadron Collider}

\author{Andrea Dainese\,\footnote{andrea.dainese@pd.infn.it}}

\address{Universit\`a degli Studi di Padova and INFN, 
         via Marzolo 8, 35131 Padova, Italy}

\begin{abstract}
With the acceleration of lead nuclei in the LHC,
heavy-ion physics will enter a new energy domain. One of the main 
novelties introduced by the 30-fold energy-jump from RHIC to the
LHC is the abundant heavy-quark production. 
After discussing a few examples of 
physics issues that can be addressed using heavy quarks, we
present a selection of results on the expected experimental 
capability of ALICE, the dedicated heavy-ion experiment at the LHC,
in the open-heavy-flavour sector. 
\end{abstract}


\section{Introduction: novel aspects of heavy-ion physics at the LHC}
\label{intro}

The nucleon--nucleon c.m.s. energy for \PbPb collisions
at the LHC, $\sqrtsNN=5.5~\tev$, will exceed that available at RHIC by 
a factor about 30, opening up a new domain for the study of 
strongly-interacting matter in conditions of high temperature and energy 
density (QCD medium). 
Particle production at the LHC will present quantitative and qualitative 
new features, as discussed in the following.



{\it Hard processes} should contribute significantly to the total 
cross section. The mechanism of energy loss 
due to medium-induced gluon radiation allows to use the energetic
partons produced in initial hard-scattering processes as probes 
to collect information on the opacity and density of the medium itself. 
At the LHC, the set of available probes will be extended both 
quantitatively and qualitatively. In fact, hard (light)quarks and 
gluons will be produced 
with high rates up to very large transverse momentum ($\pt$). 
Additionally, charm and beauty quarks, which, due to their masses, 
would show different attenuation patterns 
(see section~\ref{pheno}), will become 
available for detailed measurements, since their production cross 
sections are expected to increase by factors 10 and 100, respectively,
from RHIC to the LHC~\cite{ramona}. 

{\it High-density parton distributions} are expected to dominate particle 
production. The density of low momentum-fraction, $x$, 
gluons in the two colliding nuclei is expected to be 
close to saturation of the available phase space, so as to produce significant 
recombination effects. As an example, at central rapidity, 
low-$\pt$ c (b) quarks will be produced by partons
with
$x_1\simeq x_2\gsim 2\,m_{\rm c}(m_{\rm b})/\sqrtsNN\simeq 
2.4(10)~\gev/5500~\gev\simeq 4(16)\times 10^{-4}$. 
In section~\ref{pheno} we will discuss how gluon fusions 
in this $x$ region are expected to affect $\ccbar$ (and, to a lower
extent, $\bbbar$) production, 
not only in \pPb and \PbPb collisions (nuclear shadowing), 
but possibly even in pp collisions.

\section{Heavy-quark phenomenology from pp to nucleus--nucleus collisions}
\label{pheno}

Heavy-quark pairs, $\rm Q\overline Q$, are produced in partonic scatterings 
with large virtuality (momentum transfer) $Q\gsim 2\,m_{\rm Q}$. 
Therefore, the 
production cross sections in \NN~(NN) collisions can be calculated 
in the framework of collinearly factorized perturbative QCD (pQCD). 
The differential $\rm Q\overline{Q}$ cross section is written as:
\begin{eqnarray}
 \d\sigma^{{\rm NN\to Q\overline{Q}}X}
(\sqrtsNN,m_{\rm Q},\mu_{\rm F}^2,\mu_{\rm R}^2) 
&=&
   \sum_{i,j=\rm q,\overline q,g} 
   f_i(x_1,\mu_{\rm F}^2)\,\otimes\, f_j(x_2,\mu_{\rm F}^2)\,\otimes\nonumber\\
&& \d\hat\sigma^{ij\to {\rm Q \overline Q}\{k\}}
(\alpha_{\rm s}(\mu_{\rm R}^2),\mu_{\rm F}^2,m_{\rm Q},x_1x_2s_{\scriptscriptstyle{\rm NN}}),
\label{sigQQ}
\end{eqnarray}
where the partonic $\d\hat \sigma^{ij\to {\rm Q \overline Q}\{k\} }$ 
is calculable as a power series in the strong
coupling $\alpha_{\rm s}$, 
which depends on the renormalization scale 
$\mu_{\rm R}$;
currently, calculations are performed up to next-to-leading order (NLO),
$\mathcal{O}(\alpha_{\rm s}^3)$. 
The nucleon Parton Distribution Function (PDF) 
for the parton of type $i$ at momentum fraction $x_1$
and factorization scale $\mu_{\rm F}$, 
which can be interpreted as the virtuality 
of the hard process, is denoted by $f_i(x_1,\mu_{\rm F}^2)$.

\begin{table}
\caption{Expected $\rm Q\overline Q$ yields at the LHC, 
         from NLO pQCD
         (parameters in the text)~\cite{notehvq}. For \pPb and \mbox{Pb--Pb}, 
         PDF shadowing is included and $N_{\rm coll}$ scaling is applied.}
\label{tab:xsec}
\begin{center}
\begin{tabular}{ccccc}
\hline
colliding system & $\sqrtsNN$ & centrality & $N^{\rm c\overline{c}}$/event  &  $N^{\rm b\overline{b}}$/event  \\
\hline
pp & $14~\tev$ & minimum bias  &  0.16 & 0.0072 \\
\pPb & $8.8~\tev$ & minimum bias & 0.78 & 0.029 \\
\PbPb & $5.5~\tev$ & central (0--5\% $\sigma^{\rm tot}$) & 115 & 4.6 \\
\hline
\end{tabular}
\end{center}
\end{table}

The expected yields in pp collisions at $\sqrt{s}=14~\tev$
are reported in the first line of Table~\ref{tab:xsec}.
These numbers are obtained at NLO
using the MNR program~\cite{hvqmnr} 
with $m_{\rm c}=1.2~\gev$ and $\mu_F=\mu_R=2\,m_{\rm c}$
for charm and $m_{\rm b}=4.75~\gev$ and $\mu_F=\mu_R=\,m_{\rm b}$ for beauty;
the PDF set is CTEQ~4M~\cite{cteq4}.
The predicted yields have large uncertainties, of about a factor 2,
estimated by varying the values of the masses and of the scales
(much smaller uncertainties, $\approx 20\%$, 
arise from the indetermination in the PDFs)~\cite{yrhvq,notehvq}. 
The theoretical uncertainty band for the D-meson cross section as a function 
of $\pt$ will be shown in section~\ref{exp}, along with the expected 
sensitivity of the ALICE experiment~\cite{alicePPR}.

\begin{figure}[!t]
\begin{center}
\includegraphics[width=0.317\textwidth]{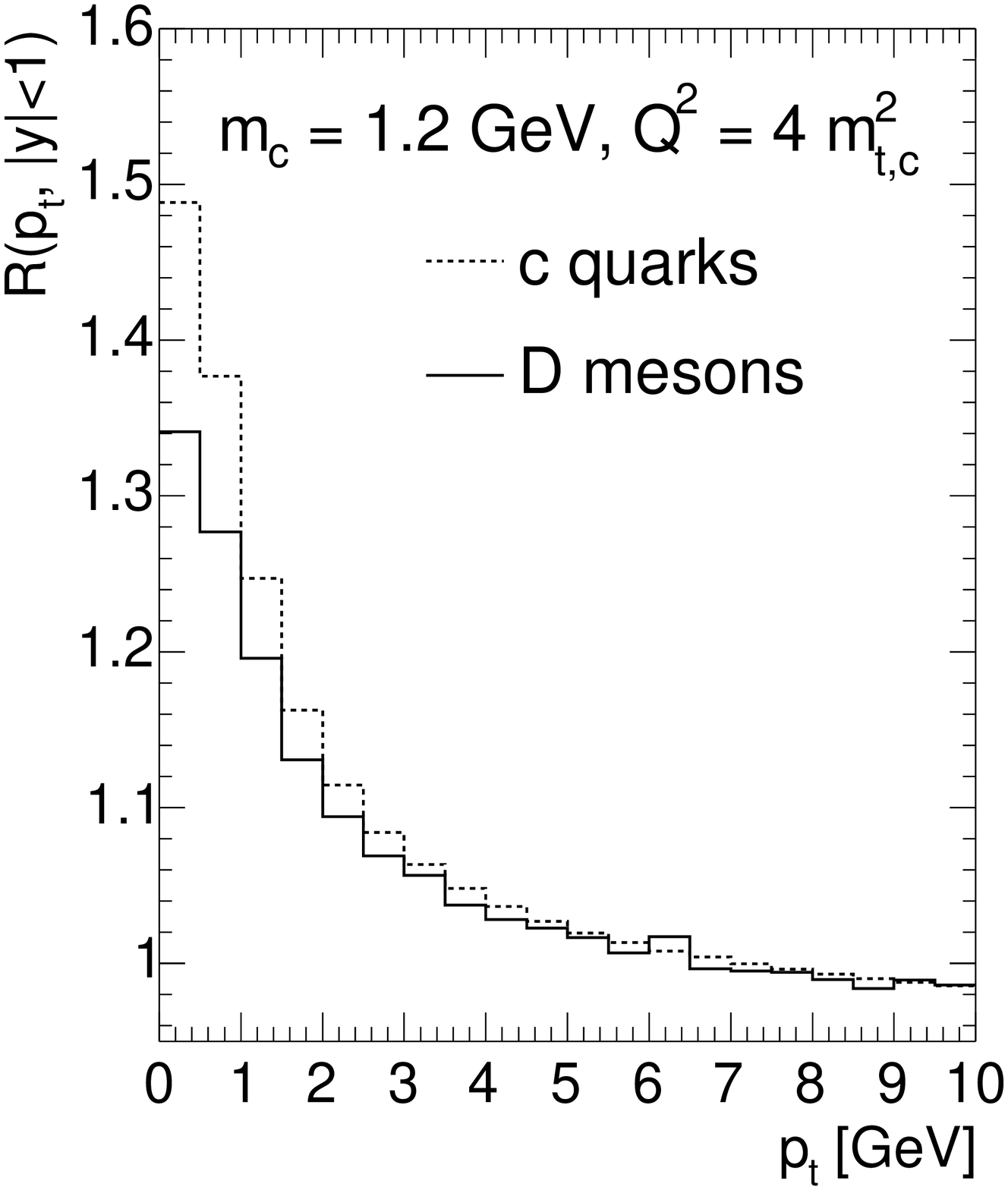}
\includegraphics[width=0.325\textwidth]{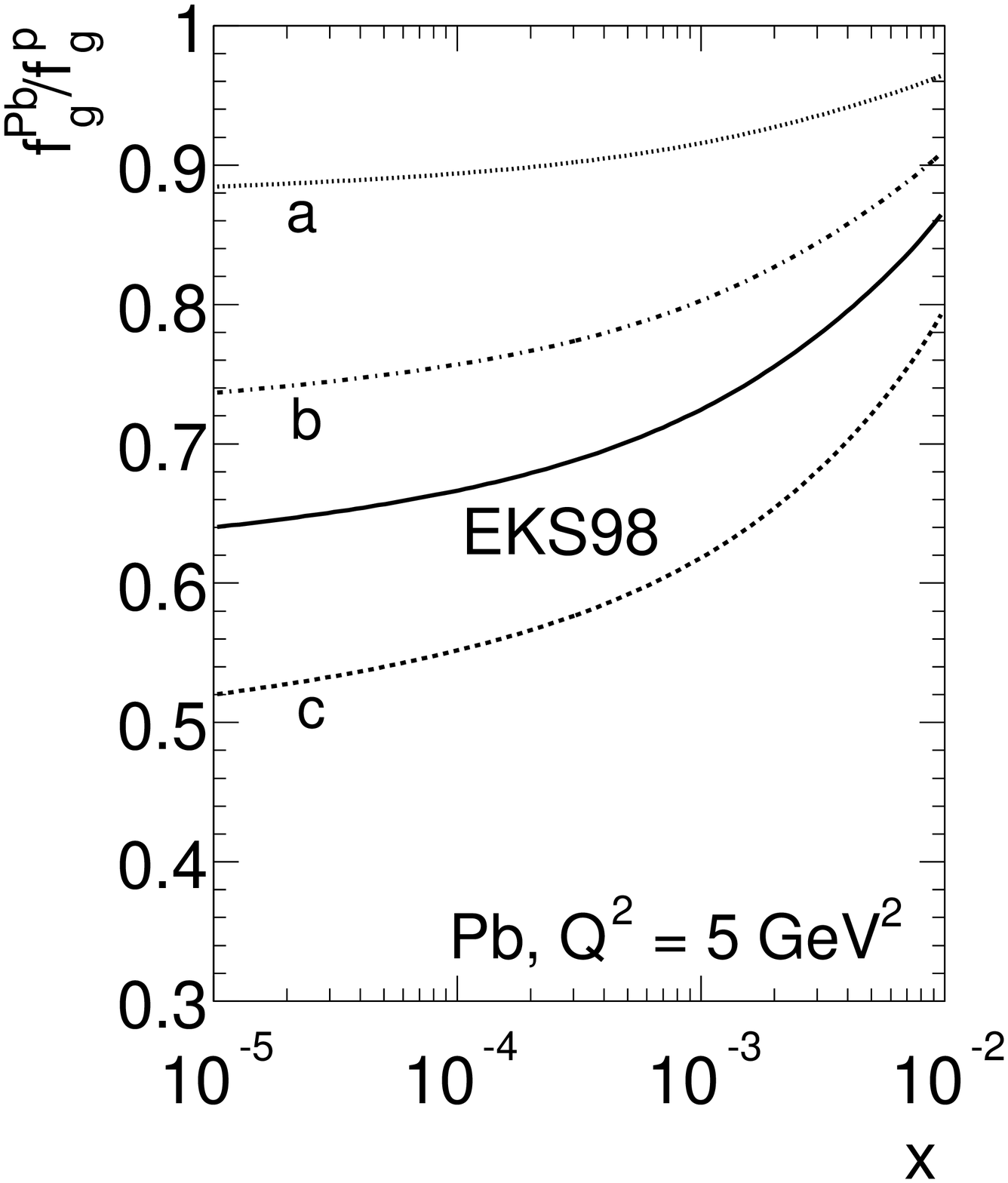}
\includegraphics[width=0.317\textwidth]{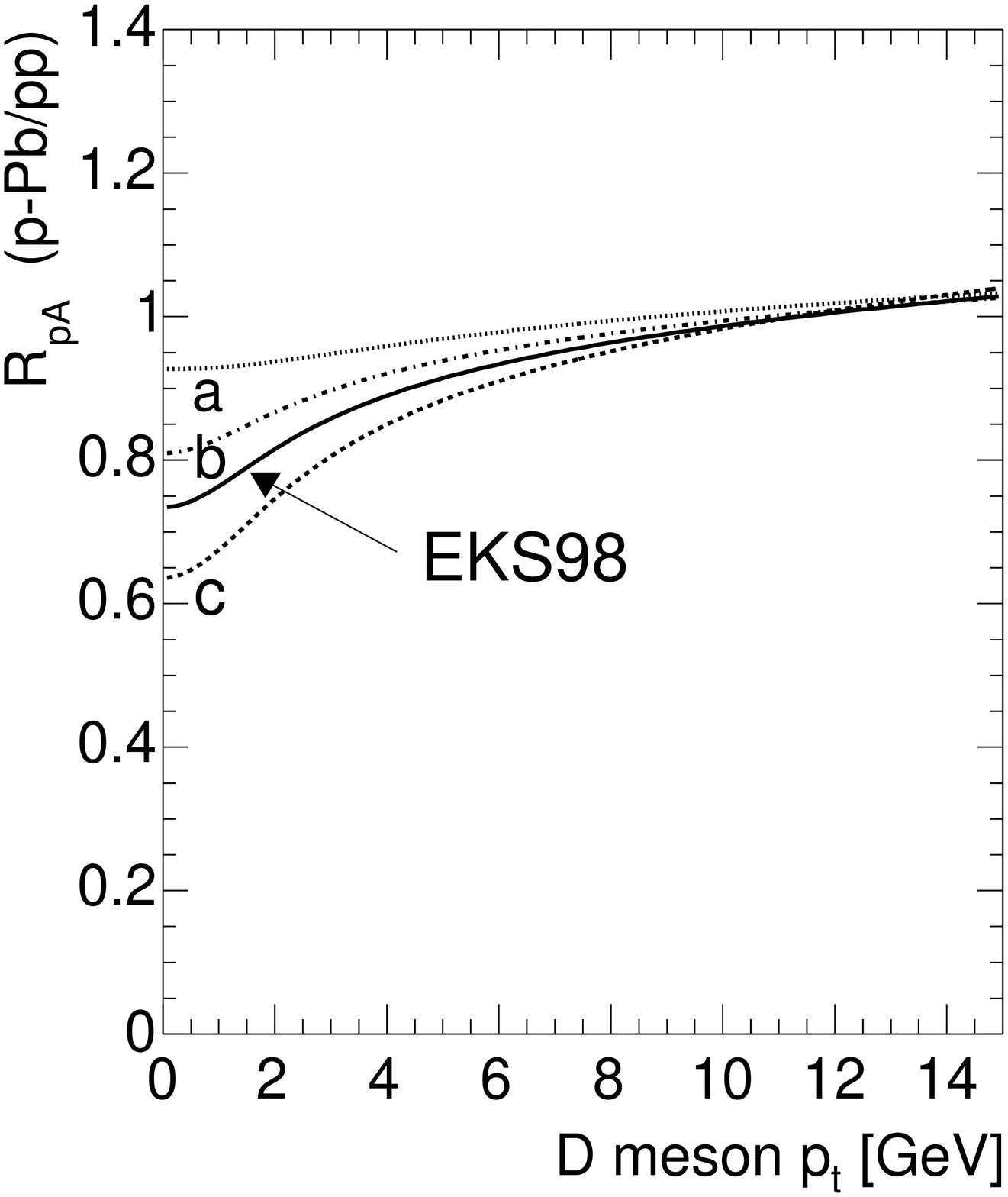}
\caption{Left: enhancement due to nonlinear gluon evolution for c quarks 
         and D mesons in pp collisions at $\sqrt{s}=14~\tev$
         \cite{dvbek}. Centre: modification of the gluon PDF 
         in a Pb nucleus at $Q^2\simeq 4\,m_{\rm c}^2$. 
         Right: corresponding
         $R_{\rm pA}^{\rm D}$ in p--Pb at $\sqrtsNN=8.8~\tev$.}
\label{fig:smallx}
\end{center}
\end{figure}

As aforementioned, in the $x$ range relevant for $\QQbar$ production, 
the PDFs will be close to phase-space saturation and, already in the case 
of pp collisions, there will be important gluon-fusion effects 
($\rm gg\to g$). 
These can be accounted for in the PDF scale-evolution equations 
by adding to the standard linear DGLAP term a negative nonlinear 
(quadratic) term (see~\cite{ehkqs} and references therein):
\begin{equation}
\partial f_{\rm g}(x,Q^2) \Big/ \partial \log Q^2= \Big[{\rm DGLAP~term~of~}
\mathcal{O}(f_{\rm g})\Big] - 
\Big[{\rm term~of~}\mathcal{O}(f_{\rm g}^2)\Big]\,.  
\end{equation}
The nonlinear term, currently calculated only at LO, 
`slows down' the $Q^2$ evolution at given $x$. It has been shown~\cite{ehkqs} 
that, for $x\lsim 10^{-2}$, it allows to have a higher gluon density 
at small $Q^2$ ($\lsim 10~\gev^2$), 
with respect to that obtained with DGLAP terms only, and to maintain at the 
same time a good fit of the proton structure function data from HERA.  
A higher gluon PDF would imply an enhancement, w.r.t. to DGLAP-based 
calculations, of $\ccbar$ production at low $\pt$ at LHC energy~\cite{ekv}. 
Figure~\ref{fig:smallx} (left) shows, as a function of $\pt$, the 
enhancement at the c-quark and at the D-meson level,
for $m_{\rm c}=1.2~\gev$ and $\mu_{\rm F}^2=\mu_{\rm R}^2=Q^2=4\,m^2_{\rm t,c}
\equiv 4\,(m_{\rm c}^2+\pt^2)$~\cite{dvbek}. 
Here, as well as for the other results presented in the following, 
the hadronization of heavy quarks is performed using the string fragmentation
model implemented in PYTHIA~\cite{pythia}.
The enhancement survives fragmentation and it is of about 30\% for D-meson 
$\pt\to 0$,
even in this `pessimistic' case where relatively-large $Q^2$ values are
considered (it should be noted, however, that this is a LO result and
the effect might be smaller at NLO). 
In section~\ref{exp} we will discuss a possible 
strategy to detect the enhancement in pp collisions at the LHC.

For hard processes, in the absence of nuclear 
and medium effects, a \AA~(or \mbox{p--nucleus}) collision 
would behave as a superposition of independent NN collisions. 
The charm and beauty differential 
yields would then scale from pp to AA (or pA) 
proportionally to the number $N_{\rm coll}$ 
of inelastic NN collisions (binary scaling):
\begin{equation} 
\d^2 N^{\scriptstyle\rm Q}_{\rm AA(pA)}/\d\pt\d y =
N_{\rm coll}\times\d^2 N^{\scriptstyle\rm Q}_{\rm pp}/\d\pt\d y\,.
\end{equation}
Binary scaling is, indeed, expected to break down due to both initial-state 
effects, such as nuclear shadowing of the PDFs, and final-state effects, 
such as parton energy loss in the medium formed in \AA~collisions. As a 
consequence,
as we will detail in the following, heavy quarks are important tools 
to probe and investigate these effects.

In Table~\ref{tab:xsec} we report the $\ccbar$ and $\bbbar$ yields 
in \pPb and \PbPb collisions calculated including in the NLO pQCD
calculation the EKS98 parameterization~\cite{eks} of the PDFs nuclear 
modification $f_i^{\rm Pb}(x,Q^2)/f_i^{\rm p}(x,Q^2)$, 
shown in Fig.~\ref{fig:smallx} (centre) for $Q^2=5~\gev^2$, 
and applying binary scaling~\cite{notehvq}. The charm (beauty) 
cross-section reduction induced by shadowing is about 35\% (20\%) in 
\PbPb and 15\% (10\%) in \mbox{p--Pb}. There is a significant 
uncertainty on the strength of shadowing in the small-$x$ region and 
some models predict much larger suppression than EKS98 
(see~\cite{yrpA} for a review). The comparison of $Q\overline Q$ production in 
pp and \pPb collisions (where final-state effects are not present) 
is regarded as a sensitive tool
to probe nuclear PDFs at the LHC.
The ratio of invariant-mass spectra of di-leptons from heavy-quark decays 
in \pPb and pp collisions would measure 
the nuclear modification $f_{\rm g}^{\rm Pb}/f_{\rm g}^{\rm p}$~\cite{yrpA}. 
Another promising observable in this respect is the nuclear modification 
factor of the D-meson $\pt$ distribution:
\begin{equation}
R^{\rm D}_{\rm pA(AA)}(\pt)=
{1\over N_{\rm coll}} \times 
{\d^2 N^{\rm D}_{\rm pA(AA)}/\d\pt\d y \over 
\d^2 N^{\rm D}_{\rm pp}/\d\pt\d y}\,.
\end{equation}
We note that at the LHC, the pp, \pPb and \PbPb runs will have different 
$\sqrtsNN$ values; however, pQCD can be used to
extrapolate the measured cross sections between different 
energies~\cite{yrhvq} and, 
thus, calculate the $R_{\rm pA(AA)}$ ratios.  
In Fig.~\ref{fig:smallx} (right) we show the sensitivity of 
$R^{\rm D}_{\rm pA}$ to 
different shadowing scenarios, obtained by varying the modification of the 
PDFs (shown for gluons in the central panel of the same figure).

\begin{figure}[!t]
  \begin{center}
  \begin{tabular}{lc}
  \begin{minipage}{0.41\linewidth}
  \includegraphics[width=1.4\textwidth]{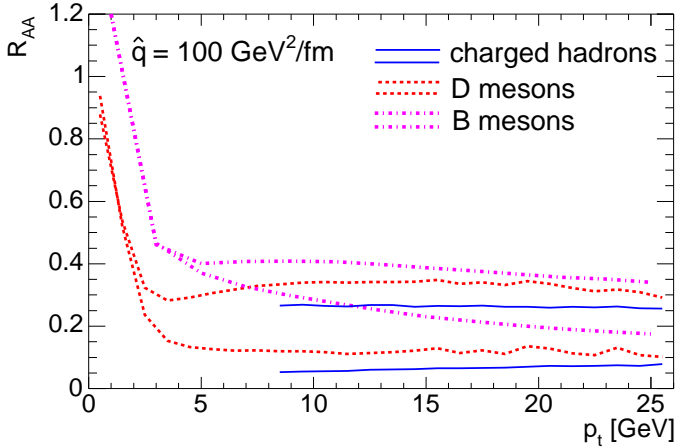}
  \end{minipage}
    &
  \begin{minipage}{0.53\linewidth}
  \caption{Nuclear modification factors for charged hadrons~\cite{pqm}, 
         D mesons ($m_{\rm c}=1.2~\gev$) and B mesons 
         ($m_{\rm b}=4.8~\gev$)~\cite{adsw}
         in central (0--10\% $\sigma^{\rm tot}$) \PbPb collisions relative 
         to binary scaling from pp collisions, at $\sqrtsNN=5.5~\tev$.
         The D-meson result does not include feed-down from ${\rm B\to D}+X$
         decays, which is, however, expected to be rather small 
         ($\sim 5\%$)~\cite{thesis}.}
  \end{minipage}
  \end{tabular}
  \label{fig:RAA}
  \end{center}
\end{figure}


Experiments at RHIC have shown that the nuclear modification factor $\RAA$ 
is an effective tool for the study of the interaction of the hard partons 
with the medium produced in nucleus--nucleus collisions.
Heavy-quark medium-induced quenching is one of the most captivating 
topics to be 
addressed in \PbPb collisions at the LHC, where both c and b quarks
will be produced with high rates (see Table~\ref{tab:xsec}). Due to the 
QCD nature of parton energy loss, quarks are predicted to lose less
energy than gluons (that have a higher colour charge) and, in addition, 
the `dead-cone effect' is expected to reduce the energy loss of massive 
quarks~\cite{dk,asw}. Therefore, one should observe a pattern 
of gradually decreasing $\RAA$ suppression when going from gluon-originated
light-flavour hadrons ($h$) to D and to B mesons: 
$\RAA^h\lsim\RAA^{\rm D}\lsim\RAA^{\rm B}$. 
In Fig.~\ref{fig:RAA} we report recent 
estimates of these quantities for central \PbPb collisions at the 
LHC~\cite{pqm,adsw}, obtained in the framework of a model~\cite{pqm} 
where energy loss is simulated in a parton-by-parton approach combining  
the BDMPS `quenching weights'~\cite{sw} and a Glauber-model-based definition 
of the in-medium parton path length. For c and b quarks,
the quenching weights were specifically calculated
using the formalism developed in~\cite{asw}. 
The BDMPS transport coefficient (a measure of the medium density) 
at the LHC was set to the value $\hat{q}=100~\gev^2/\fm$, estimated on the 
basis of the analysis of RHIC data performed in~\cite{pqm}. 
The results are plotted as bands that represent the theoretical 
uncertainty~\cite{pqm,adsw}. For charged hadrons, $\RAA$ is given for 
$\pt\gsim 8~\gev$ because at lower $\pt$ the soft-particle component 
produced from the radiated gluons, which is not implemented 
in the model, might strongly contribute in shaping the nuclear modification 
factor. Whereas for D and B mesons $\RAA$ can be calculated down to $\pt=0$
by assigning a `thermal' transverse momentum (according to 
$\d N/\d m_{\rm t} \propto m_{\rm t}\exp(-m_{t}/T)$, $T=300~\mev$) 
to the c and b quarks that lose most of their initial energy in the 
medium~\cite{adsw}. The presence of this thermalized component determines 
the rise of $\RAA^{\rm D,B}$ at $\pt\to 0$. It should be mentioned here that, 
in the low $\pt$ region, the hadronization of heavy quarks in nucleus--nucleus 
collisions is likely to happen inside the medium via parton 
recombination~\cite{molnar},
thus producing deviations from the simple pattern obtained from the 
aforementioned thermalization assumption.

\section{Looking for heavy quarks at the LHC: tools, techniques, performance}
\label{exp}

Three experiments will participate in the LHC heavy-ion program:
ALICE, the dedicated heavy-ion experiment~\cite{alicePPR}; CMS, with a
strong heavy-ion program~\cite{cms}; (most probably) ATLAS, which 
recently expressed interest in participating~\cite{atlas}. The three 
detectors have different features and design requirements, but all of them 
are expected to have excellent capabilities for heavy-flavour measurements.
Their complementarity will provide
a very broad coverage in terms of phase-space, decay channels and observables.

Experimentally, the two key elements for a rich heavy-flavour program are:
tracking/vertexing and particle identification (PID).

\begin{figure}[!t]
\begin{center}
\includegraphics[width=.75\textwidth]{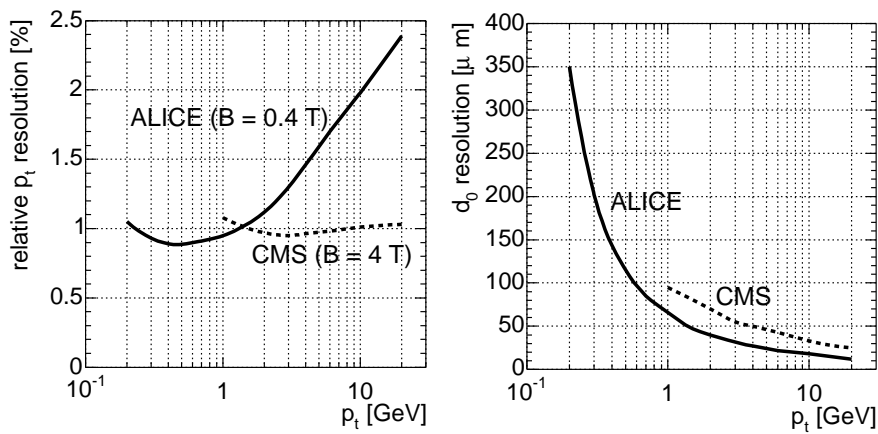}
\caption{Transverse momentum (left) and transverse track impact parameter
         (right) resolutions for the ALICE and CMS detectors in \PbPb
         collisions.}
\label{fig:ptd0}
\end{center}
\end{figure}

Open charm and beauty mesons have typical life-times of few hundred microns
($c\tau$ values are about $125$--$300~\mum$ for D mesons and $500~\mum$ for B
mesons) and the most direct detection strategy is the identification of single
tracks or vertices that are displaced from the interaction vertex.
The detector capability to perform this task is
determined by the impact parameter\footnote{We define as impact 
parameter the distance of closest approach to the interaction vertex 
of the track projection in the plane transverse to the beam direction.}
($d_0$) resolution. All experiments will be equipped with 
high position-resolution silicon-detector layers, including pixels, for
precise tracking and impact parameter measurement also in the 
high-multiplicity environment of central \PbPb collisions.
Tracking is done in the central (pseudo)rapidity region:
$|\eta|<0.9$ for ALICE and $|\eta|\lsim 1.5$ for CMS and ATLAS.  
In Fig.~\ref{fig:ptd0} we show the $d_0$ resolution of ALICE and 
CMS\footnote{ATLAS may perform similarly to ALICE,
but systematic \PbPb studies 
are still in progress.}, along
with the $\pt$ resolution, which is another important ingredient for 
heavy-flavour measurements (e.g. it determines the invariant mass resolution).
The main difference between the two 
experiments is given by the magnetic field values: ALICE (0.4~T) has a very 
low $\pt$ cutoff of $0.2~\gev$, while CMS (4~T) has a higher cutoff of $1~\gev$
but better $\pt$ resolution at high $\pt$. The $d_0$ resolutions are 
quite similar and better than $50~\mum$ for $\pt\gsim 1.5$--$3~\gev$.

Both lepton and hadron identification are important for heavy-flavour 
detection. D and B mesons have relatively-large branching ratios (BR) in the 
semi-leptonic channels, $\simeq 10\%$ to electrons and $\simeq 10\%$ to muons,
and inclusive cross-section measurements can be performed via single leptons 
or di-leptons. ALICE can identify electrons with $\pt>1~\gev$ and 
$|\eta|<0.9$, via transition radiation and $\dEdx$ measurements, and muons 
in the forward region, $2.5<\eta<4$, which allows a very low $\pt$ cutoff 
of $1~\gev$. CMS and ATLAS have 
a broad pseudorapidity coverage for muons, $|\eta|<2.4$ and $|\eta|<2.7$,
respectively, but they have a quite-high $\pt$ cutoff of 
$\approx 4~\gev$. Both CMS and ATLAS have high-resolution 
electro-magnetic calorimeters that can be used to identify 
electrons, although performance studies for heavy-ion collisions have not
been carried out yet. Semi-leptonic inclusive measurements do not
provide direct information on the D(B)-meson $\pt$ distribution, especially
at low $\pt$, because of the weak correlation between the lepton and meson
momenta. Therefore, for charm in particular, the reconstruction of exclusive
(hadronic) decays is preferable. In this case, 
in a high-multiplicity environment, 
hadron identification allows a more effective rejection
of the large combinatorial background in the low-$\pt$ region.
ALICE disposes of $\rm \pi/K/p$ separation via $\dEdx$ and time-of-flight 
measurement for $p<3$--$4~\gev$ and $|\eta|<0.9$.

In the following we present results on the expected performance for the 
detection of D and B mesons in ALICE (CMS and ATLAS studies are described 
in~\cite{yrhvq,petrushankoklay}). 
The $\ccbar$ and $\bbbar$ cross 
sections from Table~\ref{tab:xsec} are used and the charged-particles
rapidity density of a central ($0$--$5\%~\sigma^{\rm tot}$) 
\PbPb collision is assumed to be 
$\dNdy\simeq 6000$. The results, based on realistic detector simulations, 
are given for samples of $10^9$ pp events
(about 9 months of data taking) and of $10^7$ central \PbPb events
(about 1 month of data taking).

\paragraph{Charm reconstruction in ALICE.}

One of the most promising channels for open charm detection is the 
$\rm D^0 \to K^-\pi^+$ decay (and charge conjugate) that 
has a BR of $3.8\%$.
The expected yields (${\rm BR}\times\d N/\d y$ at $y=0$), 
in central \PbPb ($0$--$5\%~\sigma^{\rm tot}$) at $\sqrtsNN=5.5~{\rm TeV}$ and 
in pp collisions at $\sqrt{s}=14~{\rm TeV}$ are $5.3\times 10^{-1}$ and 
$7.5\times 10^{-4}$ per event, respectively.

The main feature of this decay topology is the presence of two tracks with 
impact parameters $d_0\sim 100~\mum$. The detection strategy~\cite{D0jpg} 
to cope with the large combinatorial background from the underlying event 
is based on the selection of displaced-vertex topologies, i.e. two tracks with 
large impact parameters and good alignment between the $\rm D^0$ momentum 
and flight-line, and on invariant-mass analysis to extract the signal 
yield.
This strategy was optimized separately for pp and \PbPb~collisions, as a 
function of the $\rm D^0$ transverse momentum, and statistical and 
systematic errors were estimated~\cite{thesis}. 

\begin{figure}[!t]
\begin{center}
\includegraphics[width=0.4\textwidth]{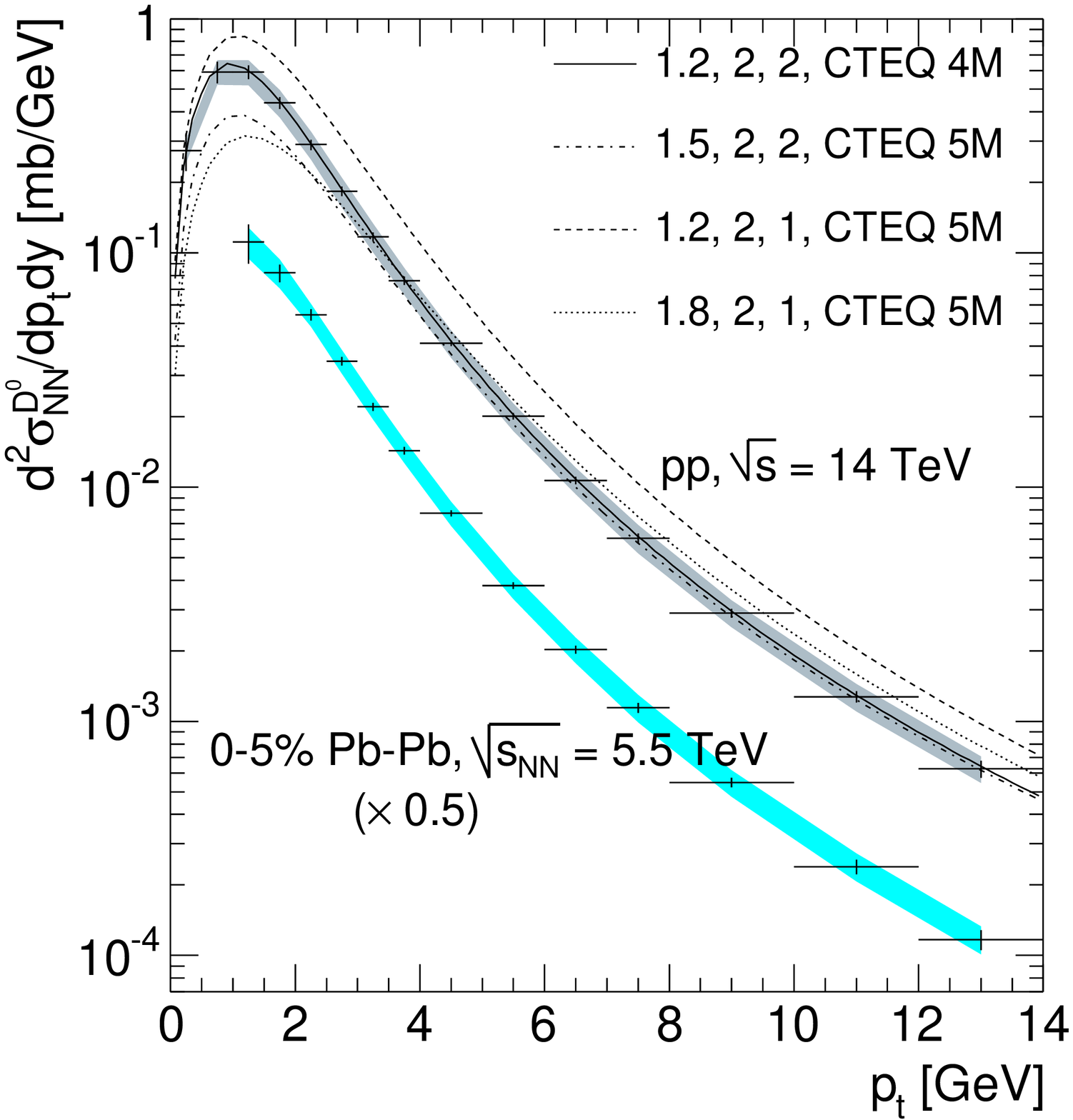}
\includegraphics[width=0.4\textwidth]{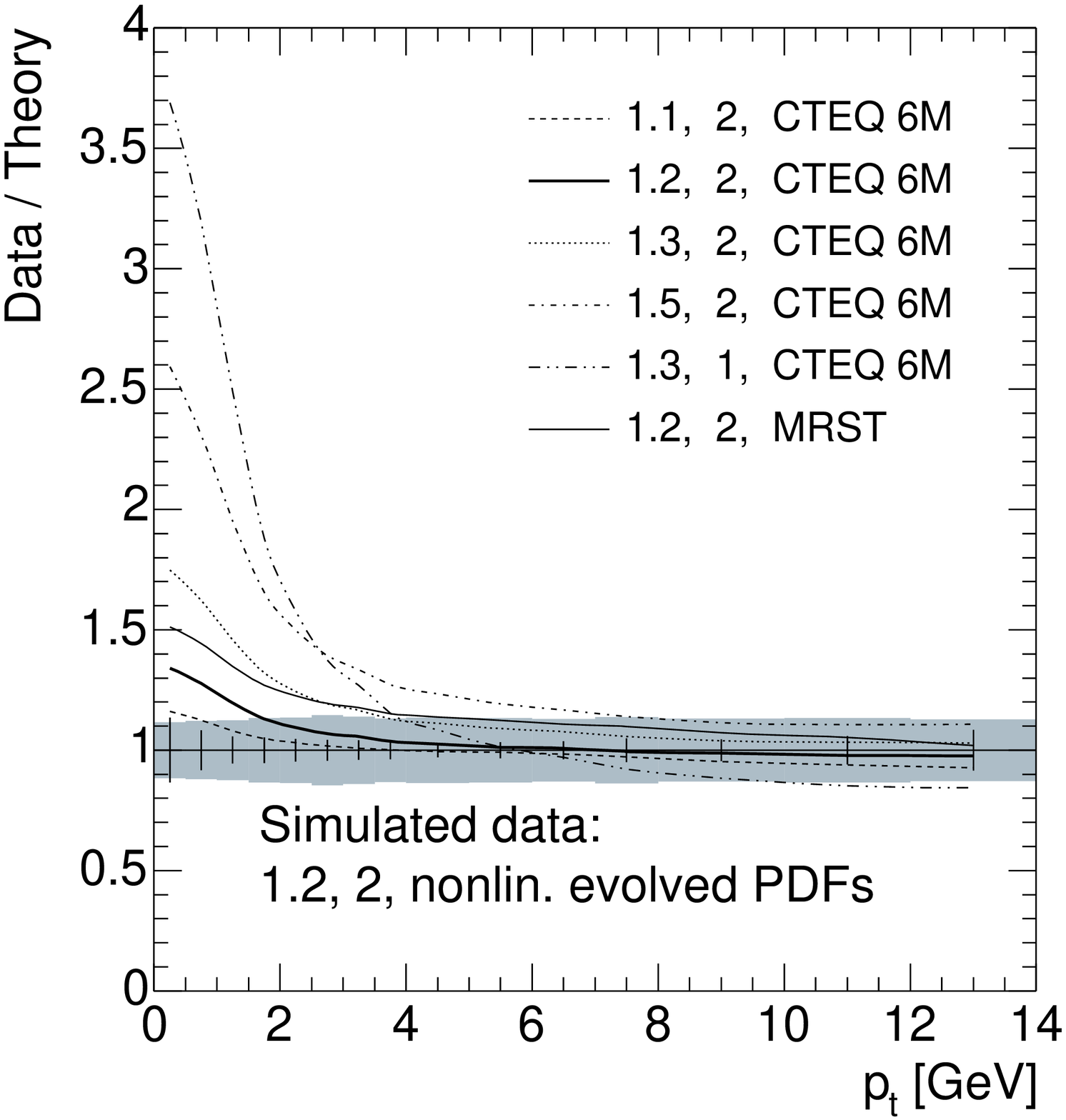}
\caption{Left: $\Dz$ production cross section
             vs. $\pt$, as it can be measured by ALICE in pp
             ($10^9$ events) and in central \PbPb ($10^7$ events) collisions; 
             statistical (bars) and $\pt$-dependent 
             systematic errors (band) are shown; a normalization error of 5\% 
             in pp and 11\% in \PbPb is not shown; 
             pQCD predictions for different sets of 
             parameters ($m_{\rm c}$~[GeV], 
             $\mu_F/m_{\rm t,c}$, $\mu_R/m_{\rm t,c}$, PDF set) 
             are also reported for the pp case.
         Right: ratios of simulated ALICE $\Dz$ data to pQCD curves 
             (parameters: $m_{\rm c}$~[GeV], 
             $Q/m_{\rm t,c}=\mu_F/m_{\rm t,c}=\mu_R/m_{\rm t,c}$, 
             PDF set); 
          the data contain the 
         enhancement due to nonlinear gluon evolution while the theory 
         curves do not~\cite{dvbek}.}
\label{fig:D0pt}
\end{center}
\end{figure}

Figure~\ref{fig:D0pt} (left) shows the expected
sensitivity of ALICE for the measurement of the $\Dz$ $\pt$-differential 
cross section in pp collisions, compared to the pQCD calculations uncertainty
that we mentioned in section~\ref{pheno}, and in central 
(0--5\% $\sigma^{\rm tot}$) \PbPb collisions.
The low-$\pt$ reach, provided by the moderate ALICE magnetic field (0.4~T) 
and the $\rm K/\pi$ separation via time-of-flight, 
allows to address the issue of charm `enhancement' in pp 
collisions due to nonlinear gluon evolution (section~\ref{pheno}). 
Here, the idea is that the effect ---enhancement only at low $\pt$---
cannot be 
mimicked by standard DGLAP-based pQCD just by varying the input parameters.
This is illustrated
in Fig.~\ref{fig:D0pt} (from~\cite{dvbek}), where the  data-to-theory ratio is 
plotted versus the $\Dz$ $\pt$. The data are obtained with 
$m_{\rm c}=1.2~\gev$ and $Q^2=4\,m_{\rm t,c}^2$ and they 
include the enhancement (from Fig.~\ref{fig:smallx} (left)), 
while the theory results do not: only for $m_{\rm c}\lsim 1.1~\gev$ the theory
can mimic the enhancement, but such small values are not supported 
by lower-$\sqrt{s}$ measurements.
 
\begin{figure}[!t]
\begin{center}
\includegraphics[width=0.52\textwidth]{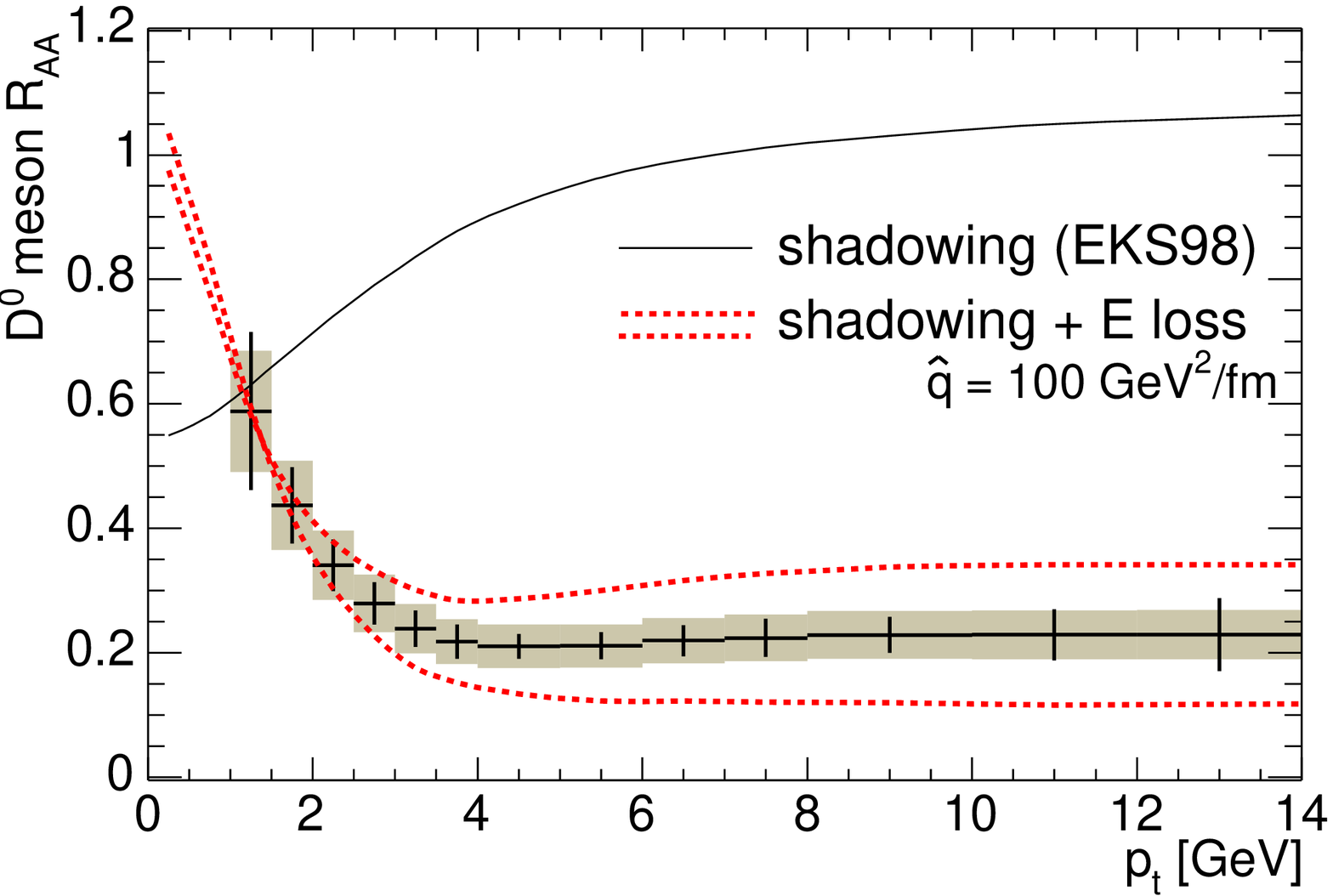}
\includegraphics[width=0.42\textwidth]{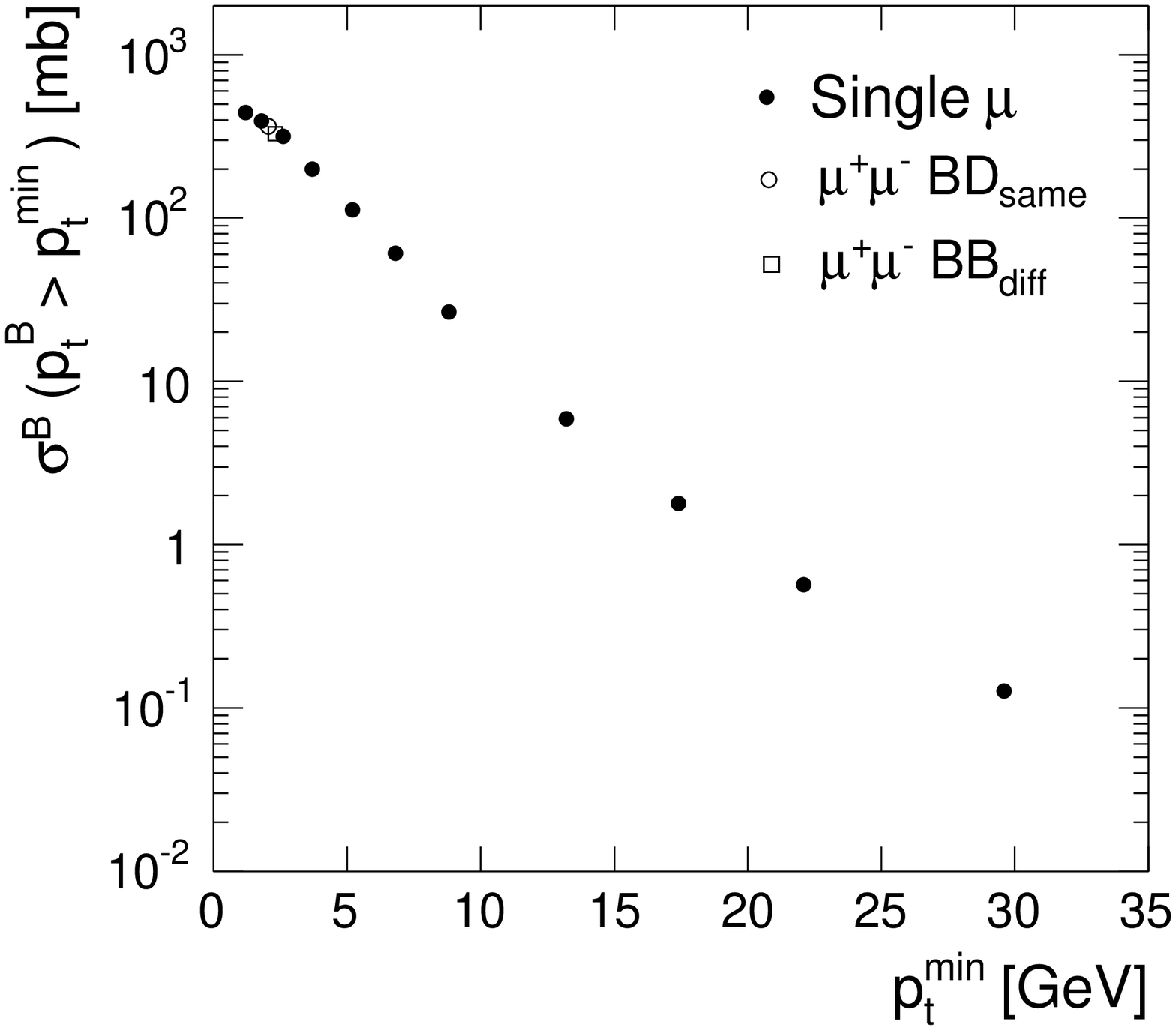}
\caption{Left: ALICE sensitivity for $\Dz$-meson $\RAA$ with $10^7$ central 
         \PbPb events and $10^9$ pp events; statistical (bars) and systematic
         (bands) errors are shown.  
         Right: B production cross section vs. $\pt^{\rm min}$ reconstructed
         by ALICE with single muons and di-muons 
         in $10^7$ central \PbPb collisions; only (very small) 
         statistical errors shown.}
\label{fig:RAADandBtomu}
\end{center}
\end{figure}

The direct measurement of the D-meson $\pt$ distribution allows a good 
sensitivity to the suppression of the nuclear modification factor due to 
c-quark energy loss in \PbPb collisions. 
Figure~\ref{fig:RAADandBtomu} (left) shows the 
estimated experimental errors on $\RAA^{\rm D}$, reported on a curve that lies 
in the middle of the D-meson band from Fig.~\ref{fig:RAA}.

Simulation studies to assess the performance for $\Dz$ reconstruction in
\pPb are in progress and the results are expected to be similar to the 
pp case~\cite{thesis}. This might allow to measure nuclear shadowing
via the $R_{\rm pA}^{\rm D}$ ratio, as discussed in section~\ref{pheno}.

\paragraph{Beauty via single electrons in ALICE.}

The expected yield (${\rm BR}\times\d N/\d y$ at $y=0$) 
for ${\rm B}\to e^{\pm}+X$ 
per central (0--$5\%~\sigma^{\rm tot}$) 
\PbPb collision at $\sqrtsNN=5.5~\tev$ is $9\times 10^{-2}$.

The main sources of background electrons are: (a) decays of D mesons; 
(b) decays of light mesons (e.g. $\rho$ and $\omega$);
(c) conversions of photons in the beam pipe or in the inner detector 
layers and (d) pions misidentified as electrons. 
Given that electrons from beauty have average 
impact parameter $d_0\simeq 500~\mum$
and a hard momentum spectrum, it is possible to 
obtain a high-purity sample with a strategy that relies on:
electron identification with a combined $\dEdx$ and transition
radiation selection, which allows to reduce the pion contamination 
by a factor $10^4$;
impact parameter cut to reject misidentified pions and electrons
from sources (b) and (c);
transverse momentum cut to reject electrons from charm decays. 
As an example, with $d_0>180~\mum$ and $\pt>2~\gev$, the expected statistics
of electrons from B decays is $5\times 10^4$ for $10^7$ central \PbPb
events, with a contamination of about 10\%, mainly given by electrons from 
charm decays~\cite{yrhvq,alicePPR2}.
The sensitivity on the extraction of the $\bbbar$ production 
cross section and of the B-meson $\pt$ distribution is currently being 
investigated.

\paragraph{Beauty via (di-)muons in ALICE.}

B production in \PbPb collisions 
can be measured also in the ALICE forward muon 
spectrometer, $2.5<\eta<4$, analyzing the single-muon $\pt$ distribution
and the opposite-sign di-muons invariant mass distribution~\cite{alicePPR2}.

The main backgrounds to the `beauty muon' signal are $\pi^\pm$, 
$\rm K^\pm$ and charm decays. The cut $\pt>1.5~\gev$ is applied to all
reconstructed muons in order to increase the signal-to-background ratio.
For the opposite-sign di-muons, the residual combinatorial background is
subtracted using the technique of event-mixing and the resulting distribution
is subdivided into two samples: the low-mass region, $M_{\mu^+\mu^-}<5~\gev$,
dominated by muons originating from a single b quark decay through
$\rm b\to c(\to \mu^+)\mu^-$ ($\rm BD_{\rm same}$), and the high-mass region,  
$5<M_{\mu^+\mu^-}<20~\gev$, dominated by $\bbbar\to\mu^+\mu^-$, with each muon
coming from a different quark in the pair ($\rm BB_{\rm diff}$). 
Both samples have a background 
from $\ccbar\to \mu^+\mu^-$ and a fit is done to extract the charm- and 
beauty-component yields. The single-muon $\pt$ distribution has three
components with different slopes: K and $\pi$, charm, and beauty decays. 
Also in this case a fit technique allows to 
extract a $\pt$ distribution of muons from B decays.
From the $\mu$-level cross sections a Monte-Carlo-based procedure is used 
to compute B-level cross sections for the data sets (low-mass $\mu^+\mu^-$, 
high-mass $\mu^+\mu^-$, 
and $\pt$-binned single-muon distribution), 
each set covering a specific B-meson $\pt>\pt^{\rm min}$ region, 
as preliminarly shown in Fig.~\ref{fig:RAADandBtomu} (right). 
Since only minimal cuts are applied, the reported statistical errors are very 
small and high-$\pt$ reach is excellent.
Systematic errors are currently under study. 

\section{Summary}
\label{summary}

We have discussed how heavy quarks, abundantly produced at LHC energies, 
will allow to address several physics issues, in 
\mbox{proton--proton},
\mbox{proton--nucleus} and \mbox{nucleus--nucleus} collisions. In particular,
they provide tools to:
\begin{itemize}
\item probe, via parton energy loss and its predicted mass dependence, 
      the high-density QCD medium formed in \PbPb collisions;  
\item probe, in pp collisions, the pQCD calculations parameters space;
\item probe the small-$x$ regime of the PDFs, 
      where saturation/recombination effects
      are expected to be important, even in pp collisions.
\end{itemize}
The excellent tracking, vertexing and particle identification performance 
of ALICE, ATLAS and CMS will allow to fully explore this rich phenomenology,
as we have shown with some specific ALICE studies on D and B meson 
measurements.

\paragraph{Acknowledgment.} The author, member of the ALICE Collaboration, 
would like to thank the ALICE off-line group, within which part of the 
results here reported have been obtained. Fruitful discussions on the 
manuscript with F.~Antinori are also acknowledged.

\vspace{.4cm}

\end{document}